\documentclass[usenatbib,fleqn]{mnras}
\usepackage{natbib,times}
\usepackage{newtxtext,newtxmath}
\usepackage[T1]{fontenc}
\usepackage{ae,aecompl}
\usepackage{graphicx}	
\usepackage{amsmath}	
\usepackage{amssymb}	
\usepackage{hyperref}
\usepackage{cleveref}
\usepackage{aas_macros}
\usepackage{textcomp}
\usepackage[utf8]{inputenc}

\usepackage{etoolbox}
\makeatletter
\patchcmd\@combinedblfloats{\box\@outputbox}{\unvbox\@outputbox}{}{\errmessage{\noexpand patch failed}}
\makeatother

\title[Type II migration]{Type II migration strikes back -- An old paradigm for planet migration in discs}

\author[C. E. Scardoni et al]{
Chiara E. Scardoni,$^{1,2}$\thanks{E-mail: ces204@cam.ac.uk}
Giovanni P. Rosotti,$^{3,2}$
Giuseppe Lodato$^{1}$ and Cathie J. Clarke$^{2}$
\\
$^{1}$Universit\`a degli Studi di Milano, Via Giovanni Celoria 16, I-20133 Milano, Italy\\
$^{2}$Institute of Astronomy, University of Cambridge, Madingley Road, Cambridge CB3 OHA, UK\\
$^{3}$Leiden Observatory, Leiden University, P.O.~Box 9513, NL-2300~RA Leiden, the Netherlands
}

\date{Accepted 2019 December 13. Received 2019 December 13; in original form 2019 November 5}

\pubyear{2019}

\begin{document}
\label{firstpage}
\pagerange{\pageref{firstpage}--\pageref{lastpage}}
\maketitle

\begin{abstract}
In this paper we analyse giant gap-opening planet migration in protoplanetary discs, focusing on the type II migration regime. According to standard type II theory, planets migrate at the same rate as the gas in the disc, as they are coupled to the disc viscous evolution; however, recent studies questioned this paradigm, suggesting that planets migrate faster than the disc material. We study the problem through 2D long-time simulations of systems consistent with type II regime, using the hydrodynamical grid code FARGO3D. Even though our simulations confirm the presence of an initial phase characterised by fast migration, they also reveal that the migration velocity slows down and eventually reaches the theoretical prediction if we allow the system to evolve for enough time. We find the same tendency to evolve towards the theoretical predictions at later times when we analyse the mass flow through the gap and the torques acting on the planet. This transient is related to the initial conditions of our (and previous) simulations, and is due to the fact that the shape of the gap has to adjust to a new profile, once the planet is set into motion. Secondly, we test whether the type II theory expectation that giant planet migration is driven by viscosity is consistent with our simulation by comparing simulations with the same viscosity and different disc mass (or viceversa). We find a good agreement with the theory, since when the discs are characterised by the same viscosity, the migration properties are the same.
\end{abstract}


\section{Introduction}
\label{sec: Introduction}

The migration of satellites in accretion discs is a fundamental process in astrophysics, both in proto-planetary discs and in discs around black holes. In the former case, for example, disc-assisted migration is a leading theory to explain the origin of hot Jupiters \citep{Lin1996} and it provides a convenient way in the solar system to explain the small mass of Mars and the asteroid belt (Grand tack scenario, \citealt{Walsh2011}). Migration tends to create compact, resonant chains, which are destroyed by dynamical instabilities after gas dissipation \citep{Izidoro2017,Liu2017}. This is in accord with the low observed number of exoplanet pairs in resonance \citep{Lissauer2011}. In the latter case instead, migration can bring super massive black holes to merge \citep{Armitage2002,Cuadra2009}, possibly overcoming the so-called final parsec problem \citep{2001ApJ...563...34M,Lodato09}. The merging leads to the emission of gravitational waves that will be observable by the future space mission LISA.

Satellite migration is not a recent topic: its seminal papers \citep{1979MNRAS.186..799L,1980ApJ...241..425G} were written several decades ago (for more recent reviews, see \citealt{KleyNelson2012,BaruteauPPVI}). Despite its age, migration is far from being a solved problem. While angular momentum conservation tells us that it is a fundamental process that (at least to some extent) \textit{must} happen as a result of satellite-disc interaction, there are still no observational, \textit{direct} proofs of its existence. In the future, thanks to high-resolution imaging of proto-planetary discs, this should be finally possible \citep{Meru2019,Nazari2019,Perez2019}, but this has not happened yet. Motivated by this lack of proof, some planet formation models simply neglected migration altogether, building planets \textit{in situ} \citep{ChiangLaughlin2013,Hansen2013}. This is attractive to bypass the extremely fast timescales of the so-called type I migration\footnote{This nomenclature is due to \citet{Ward1997}}, characteristic of low-mass (non gap-opening) planets. Type I migration is so fast that, if unimpeded, would quickly move almost all planets at the inner edge of the disc \citep{IdaLin2008}. Addressing this problem has led to a flurry of theoretical studies, involving for example the existence of planet ``traps'' in the disc capable of stopping or even reversing planet migration \citep{Hasegawa2011,Bitsch2013}. Moreover, the growing realisation that disc viscosity might be lower than previously assumed opens up many new migration regimes that are still unexplored \citep{McNally2019}. 

In this paper we are concerned however with the type II regime of massive, gap-opening planets and we will therefore discuss the issue of type I migration no further. Anecdotally, while type I migration has always been identified as problematic, traditionally type II was deemed to be the ``easy'', well-understood regime. This perceived simplicity comes from the fact that, differently from type I migration, type II does not depend on the details of planet-disc interaction. The co-orbital region, responsible for driving fast orbital migration in the type I regime, is severely depleted due to gap opening and therefore does not contribute significantly to the migration rate. As a result, as first proposed by \citet{1986LinPapaloizou}, the planet maintains a location in quasi-steady state at the centre of the gap due to the competition between the torque from the inner disc (which pushes the planet outwards) and that from the outer disc (which pushes it inwards). However, because the gas on both sides of the gap moves towards the star with the viscous velocity, the planet finds itself entrained in the viscous flow, as long as it can be easily carried by the gas. This conditions corresponds to the planet being less massive than the local disc \citep{SyerClarke1995,1999IvanovAl}.

This picture of type II migration comes from one-dimensional calculations in which the gap is a ``dam'': no material can flow through it. In reality, the picture is more complicated; it is well known that gas does flow through gaps \citep{LubowDangelo2006}. This potentially breaks the assumption behind the type II argument: if the gas can flow through the gap, it can also move at a different velocity from the planet, which therefore is not forced to move at the viscous velocity. Despite these concerns, early hydro-dynamical calculations that allowed the planet to move through the disc \citep{Nelson2000,Schafer2004,Dangelo2005} seemed to show that the migration rates were indeed compatible with the type II estimate.

More recently, however, other works \citep{2014Duffell,2015DuermannKley} have cast doubts about this picture of type II migration. A re-examination of those works showed that those simulations were actually in the regime in which the planet is more massive than the disc, and therefore it should have migrated \textit{slower} than type II. Indeed, when the disc mass was increased to satisfy the requirement that the disc be more massive than the planet, consistently \textit{faster} migration rates than type II were found. 

The goal of this paper is to revisit this problem. In particular, being a viscous process, it is a reasonable expectation that type II migration should be established on timescales comparable to the local viscous timescale. Due to the high computational cost of planet-disc interaction simulations, however, the timespan of a typical simulation is only a few hundred (thousands at best) orbits, which for typical parameters represents only a small fraction of the viscous timescale. In this paper we aim to study the long-term behaviour of the migration rate.

The paper is structured as follows: Section~\ref{sec:Disc-planet interaction} outlines the principal notions of type II migration; in Section~\ref{sec: Numerical setup description} we describe the numerical method and introduce the parameters used in our disc models; Section~\ref{sec: Standard setup analysis} is focused on the description of our standard model; we explore the effects on migration of changing the disc aspect ratio in Section~\ref{sec: Changing the value of H/R}; Section~\ref{sec: Discussion} is aimed at discussing the results and comparing them with previous works; our main conclusions are outlined in Section~\ref{sec: Conclusions}.

\section{Disc-planet interaction}
\label{sec:Disc-planet interaction}
We summarise here the main concepts behind type II migration and introduce the notation we will use in the rest of the paper.

Both gap opening and planet migration are caused by the exchange of angular momentum between the planet and the disc through their tidal interaction. The equation describing the disc evolution under the effect of viscosity, including the additional tidal torque, is:
\begin{equation}
    \frac{\partial\Sigma}{\partial t}=\frac{3}{R}\frac{\partial}{\partial R}\left[R^{1/2}\frac{\partial}{\partial R}(R^{1/2}\nu\Sigma)\right]-\frac{2}{R}\frac{\partial}{\partial R}\left(\frac{\Sigma\Lambda_{\rm T}}{\Omega}\right),
    \label{eq:disc_evolution}
\end{equation}
where $\Sigma$ is the disc surface density, $\nu$ the kinematic viscosity, $\Omega$ the orbital velocity and $\Lambda_{\rm T}$ the tidal specific torque. Rigorously, $\Lambda_{\rm T}$ should be computed as combination of the contributions provided at the Lindblad resonances \citep{1979GoldreichTremaine}; however, the so-called "impulse approximation" \citep{1979MNRAS.186..799L}, assuming that the disc-planet interaction can be described as a scattering process, leads to the correct form for the torque
\begin{equation}
    \Lambda_{\rm T}=\frac{q^2}{2}(\Omega R)^2\left(\frac{a}{\Delta_{\rm gap}}\right)^4 \mathrm{sgn}(R-a),
\end{equation}
being $q=M_{\rm p}/M_{*}$ the planet-to-star mass ratio, $a$ the planet semi-major axis and $\Delta_{\rm gap}$ the gap width.

By imposing the system angular momentum conservation, we obtain the second equation describing the system, giving the evolution of the planet semi-major axis:
\begin{equation}
    \frac{d}{dt}(M_{\rm p}\Omega_{\rm p}a^2)=-\int_{R_{\rm in}}^{R_{\rm out}}2\pi R\Lambda_{\rm T}\Sigma dR,
\end{equation}
whose right-hand side strength is given by the parameter
\begin{equation}
    B=\frac{4\pi a^2\Sigma_0}{M_{\rm p}},
    \label{eq:B}
\end{equation}
where $\Sigma_0$ would be the disc surface density at planet location if the planet were not present.

\begin{table*}
    \centering
    \caption{Parameters used in the simulations.}
    \begin{tabular}{ccccccccr}
    Simulation name &    $N_r$  &   $N_{\phi}$   &   $h_0$   &   $\sigma_0$   &   $\alpha$   &   $q$   &   $r_{\rm in}$&   $r_{\rm out}$\\
    \hline
    Standard disc &    $251$  &   $583$   &   $0.05$   &   $0.0026698$   &   $0.003$   &   $0.001$   &   $0.3$   &   $3$\\
    Thin disc &    $314$  &   $729$   &   $0.04$   &   $0.0026698$   &   $0.003$   &   $0.001$   &   $0.3$   &   $3$\\
    Thin-heavy disc &    $314$  &   $729$   &   $0.04$   &   $0.0041717$   &   $0.003$   &   $0.001$   &   $0.3$   &   $3$\\ 
    Thick disc &    $310$  &   $570$   &   $0.06$   &   $0.0026698$   &   $0.003$   &   $0.001$   &   $0.1$   &   $3$\\ 
    Thick-light disc &    $310$  &   $570$   &   $0.06$   &   $0.0018541$   &   $0.003$   &   $0.001$   &   $0.1$   &   $3$\\ 
    \end{tabular}

    \label{tab:parameters}
\end{table*}

Depending on the value of parameter $B$, we can distinguish two regimes of type II migration: the "disc dominated regime" or "classical type II migration", which occurs when the disc is more massive than the planet ($B>1$); and the "planet-dominated regime", characteristic of planets more massive than the disc ($B<1$). The general type II migration velocity can be expressed as
\begin{equation}
    u_{\rm II}=u_{r}\frac{B}{B+1},
    \label{eq:uII}
\end{equation}
where $u_{r}$ is the gas viscous radial velocity. In the disc-dominated regime, the planet migration velocity approaches the gas radial velocity; the larger the value of $B$, the more $u_{\rm II}$ is similar to $u_{r}$, and in the limit case $B\gg 1$ they have exactly the same value. In this migration regime the planet is coupled to the disc viscous evolution: the viscous accretion of the inner disc causes the inner part of the gap to change its location; as the outer gas cannot cross the gap to re-fill the inner gap edge, the planet must migrate inwards to restore the equilibrium; then also the outer disc moves viscously inward.
On the contrary, in the planet-dominated regime, the planet's inertia prevents the planet from migrating as fast as the disc material; in the limit case $B\ll 1$, the planet velocity is highly reduced, as $u_{\rm II}\sim B u_{r}$. In this scenario, the planet is almost stuck at its location and it hinders the outer disc accretion (as the outer gas cannot cross the gap), causing the disappearance of the gas internal to the planet orbit \citep{SyerClarke1995,1999IvanovAl}.

An alternative route to consider the problem is through the migration timescale, instead of the migration velocity; in this approach, \autoref{eq:uII} can be expressed as
\begin{equation}
    t_{\rm II}=t_{\nu,0}\frac{B+1}{B},
\end{equation}
where $t_{\nu, 0}$ is the local viscous timescale, i.e. the viscous timescale computed at planet location
\begin{equation}
    t_{\nu,0}\propto \frac{a^2}{\nu(a)}.
\end{equation}
In the disc-dominated regime, therefore, we expect the migration time to approach to the viscous timescale, whereas in the planet-dominated regime the migration time is expected to be much longer than the viscous timescale.

The advantage of focusing on the migration timescale instead of the migration velocity is that we can easily distinguish the migration timescale from the others fundamental timescales involved in this problem: the disc viscous timescale and the gap-opening timescale. The former is the time required by the system for its viscous evolution; it can be obtained by computing the viscous timescale at the outer disc radius
\begin{equation}
    t_{\nu, \rm disc}\propto \frac{R_{\rm out}^2}{\nu(R_{\rm out})}.
\end{equation}
The latter timescale is the amount of time required for the gas next to the planet to acquire (or lose) angular momentum through the interaction with the planet, and then to be pushed out of the gap area. It can be, therefore, evaluated as the ratio between the required change in angular momentum and the angular momentum transfer rate
\begin{equation}
    t_{\rm gap}=\left|\frac{\Delta J}{\dot{J}}\right|.
\end{equation}

\section{Numerical setup description}
\label{sec: Numerical setup description}
To study type II migration properties, we used the hydrodynamical grid code FARGO3D \citep{2016Benitez-Llambay&Masset}. We performed 2D simulations, under the assumption of geometrically thin discs, and we defined our ``standard disc'' with the same characteristics as the standard setup used by \citet{2015DuermannKley}, for comparison purposes. As natural choice for the problem geometry, we adopted a cylindrical reference frame $(r,\phi)$, with the star located at the centre. We used dimensionless units: the initial planet location $a_{0}$ is the unit length, the star mass $M_{*}$ is the unit mass and the Keplerian orbital frequency $\Omega_{\rm k}^{-1}$ at the initial planet location is the  unit time. For convenience, we will rescale the time unit in the rest of the paper in units of orbits at the initial planet location.

In the standard model, we considered a grid made of $N_{\phi}=583$ cells in the azimuthal direction (from $0$ to $2\pi$), and $N_r=251$ cells in the radial direction (from $r_{\rm in}=0.3$ to $r_{\rm out}=3.0$). The grid cells are logarithmically spaced in the radial direction; the radial resolution is chosen in order to enforce the same resolution in both directions. The equation of state was assumed to be locally isothermal and the viscosity was parameterised using the $\alpha$-prescription \citep{1973ShakuraSunyaev}
\begin{equation}
    \nu =\alpha c_{\rm s}H,
    \label{eq:alphavisc}
\end{equation}
where $\alpha=0.003$ in all the simulations performed in this work.

In the standard setup we considered a uniform aspect ratio $h_0=0.05$ at the initial planet location. In addition, we performed simulations characterised by different values for $h_0$ ($h_0=0.06,0.04$); in these cases we adjusted accordingly the number of cells to obtain the same resolution in units of scaleheight as in the standard setup resolution. Note that in the case $h_0=0.06$ we extended the inner edge of the simulated disc to $r_{\rm in}=0.1$, therefore we rescaled $N_{r}$ also considering the domain extension in radial direction.

It should be noticed that accretion of disc material by the planet is not considered in our model; however, accretion onto the planet should not produce significant effects on planet migration, as recently shown by \citet{2018Robert}.

For a summary of the parameters considered in the simulations, see \tableautorefname~\ref{tab:parameters}.

\subsection{Initial and boundary conditions}
We designed initial and boundary conditions to obtain a steady-state disc. Our objective is thus to achieve a constant mass flow $\dot{m}$ through the disc:
\begin{equation}
    \dot{m}=-2\pi r \sigma u_{\rm r}=\rm const.
    \label{eq:mdot}
\end{equation}
Since the aspect ratio is constant with radius the viscosity $\nu\propto \Omega R^2\propto R^{1/2}$ and thus, to obtain a steady disc, we need the following density profile
\begin{equation}
    \sigma(r) =\sigma_0 r^{-1/2},
    \label{eq:densityprofile}
\end{equation}
where $\sigma_0$ is the density at $a_0=1$. As a consequence, the radial velocity must be
\begin{equation}
    u_{r}(r)=-\frac{\dot{m}}{2\pi \sigma_0}r^{-1/2}.
    \label{eq:radialvelprofile}
\end{equation}

To speed up the steady state achievement, we chose an initial density profile that already matches the steady state profile (\autoref{eq:densityprofile}). Similarly, the initial condition for the radial velocity profile was implemented using \autoref{eq:radialvelprofile}. The initial azimuthal velocity was written using the prescription by \citet{2004MassetOgilive}, which includes pressure corrections in order to obtain an accurate initial centrifugal balance and avoid numerical mismatches that may cause the occurrence of waves perturbing the system equilibrium \citep{2004MassetOgilive}.

The boundary condition on the velocity at inner and outer grid edges is chosen to match \autoref{eq:radialvelprofile} also in the ghost cells. The outer boundary condition for the density was chosen to provide the disc with a constant mass inflow (\autoref{eq:mdot}), therefore $\sigma_{\rm out}=\sigma_0 r_{\rm out}^{-1/2}$. At the inner boundary we allowed the density to leave the computational domain, considering an open boundary condition.

To avoid numerical reflections at the grid boundaries, we introduced wave-killing boundary conditions for both velocity components. Using the method of \citet{2006Val-Borro}, we damped $u_r$ and $u_{\phi}$ to their azimuthal average $\langle u_{r}\rangle_{\phi}$ and $\langle u_{\phi}\rangle_{\phi}$, over a timescale equal to the local $\Omega^{-1}$. The radial extent of the cells involved in these boundary conditions (as a fraction of the whole computational domain) is $5\%$ at the inner boundary and $10\%$ at the outer boundary.

\section{Standard disc analysis}
\label{sec: Standard setup analysis}
\subsection{System preparation for planet migration}
\begin{figure}
    \centering
    \includegraphics[width=1\linewidth]{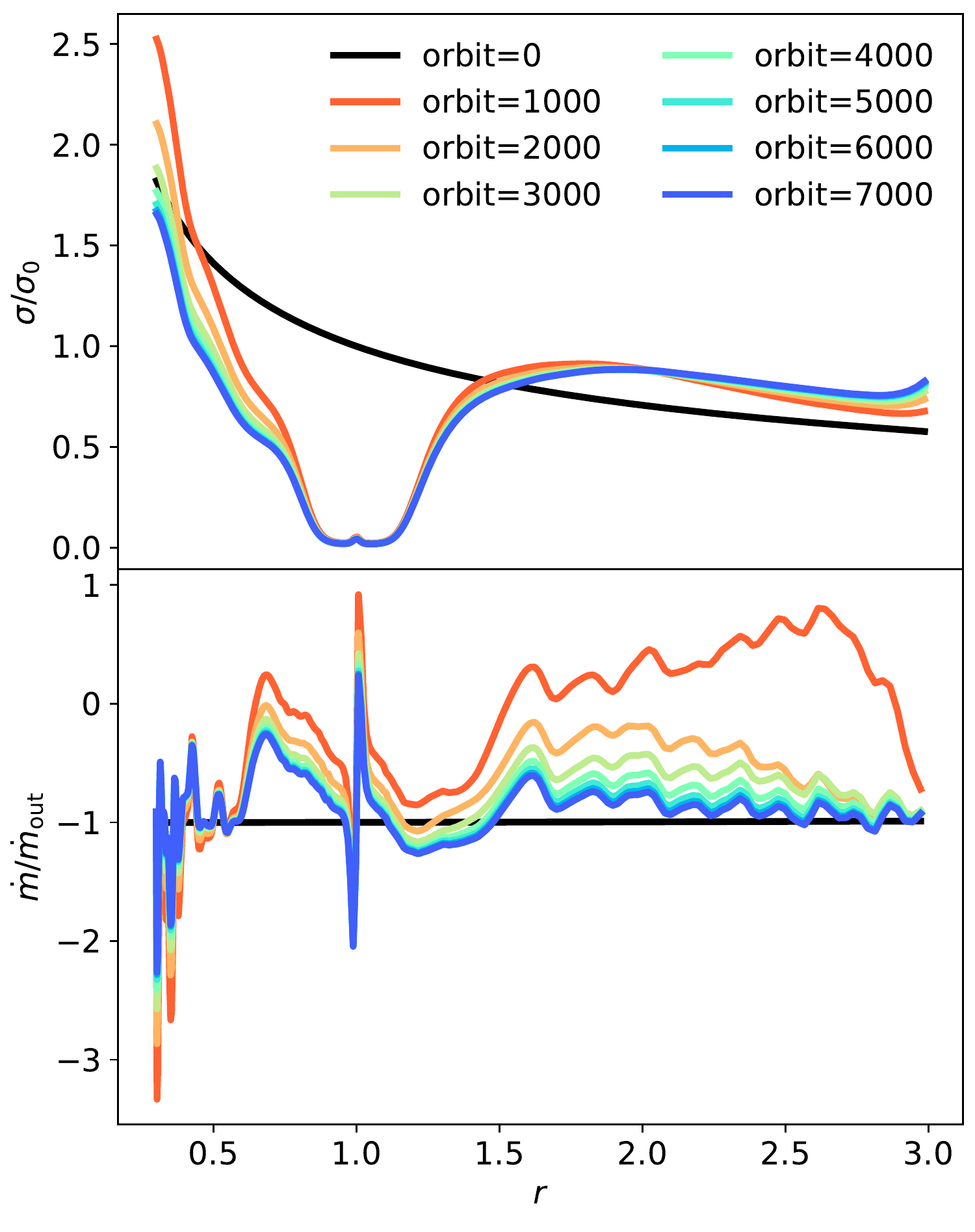}
    \caption{Standard disc: azimuthally averaged density (upper panel) and accretion rate through the disc (lower panel) before (black line) and during (coloured lines) planet insertion.}
    \label{fig:preparazione}
\end{figure}

After having defined the setup as described in section~\ref{sec: Numerical setup description}, we allowed the system to relax until reaching the steady state equilibrium ($t_1=5000$ orbits).
After $5000$ orbits, we computed the azimuthally averaged disc density (black line in the upper panel of Fig.~\ref{fig:preparazione}) and we found that it matches the steady state profile $\sigma\propto r^{-1/2}$. To confirm the steady state achievement we verified that the normalised accretion rate through the disc (black line in the lower panel of Fig.~\ref{fig:preparazione}) is constant.

Then we gradually inserted the planet, by increasing its mass from $M_{\rm p}=0$ to $M_{\rm p}=10^{-3}M_\star$ over $1000$ orbits, to minimise numerical perturbations induced by planet insertion. As the parameter $q$ is defined as the planet to star mass ratio $q=M_{\rm p}/M_{*}$, when we consider a Sun-like star, $q=10^{-3}$ means that we are considering a Jupiter-like planet. In particular, we used the tapering function $q (t)=10^{-3}\cdot 1/2 \left[1.0-\cos\left(\pi (t-t_{1})/t_{\rm p}\right)\right]$, where $t_{\rm p}=1000$ orbits.
Note that when the planet is inserted in the simulation we must smooth its gravitational potential \citep{2012Muller}
\begin{equation}
    \Phi_{\rm planet}(s)=-\frac{GM_{\rm p}}{\sqrt{s^2+\epsilon^2}},
\end{equation}
where $s=r-a$ and $\epsilon$ is the smoothing, chosen to be $\epsilon=0.6\ H$.
Furthermore, we excluded a portion of disc next to the planet in the torques computation, using the prescription by \citet{2008Crida}
\begin{equation}
    \zeta (s)=\left[\exp\left(-\frac{s/r_{\rm H}-\xi }{\xi/10}\right)+1\right]^{-1},
    \label{eq:excluderH}
\end{equation}
where $s$ is the gas distance from the planet, $r_{\rm H}$ is the Hill radius, $\xi$ is the fraction of the Hill radius excluded from the torques computation (we chose $\xi=0.8$).

Once planet insertion was completed, we evolved the system for $7000$ additional orbits to allow the disc to adapt to the planet presence and re-obtain the steady state condition.
The coloured lines in the upper panel of Fig.~\ref{fig:preparazione} illustrate the average azimuthal density evolution when the planet is kept fixed at its location (the orbit count has been reset after the complete planet insertion). As from $5000$ orbits to $7000$ orbits the density hardly evolves, we deduce that the system has approximately reached the steady state condition. The steady state achievement is confirmed by the plot in the lower panel (cf. figure 4 in \citealt{2015DuermannKley}), which shows the accretion rate (obtained integrating the mass flow over the azimuthal direction) as a function of the radius: apart from modest oscillations, at $7000$ orbits we found $\dot{m}/\dot{m}_{\rm out}=-1$ (blue curve).

\subsection{Planet migration}
At this point, the system preparation is completed, thus we let the planet migrate until it reaches the inner boundary of the grid, which happens after approximately 1650 orbits. During the evolution of the system, we evaluated the variation in the planet semi-major axis $a$ and we compared it to its theoretical value (\autoref{eq:uII}).

\begin{figure}
    \centering
    \includegraphics[width=0.9\linewidth]{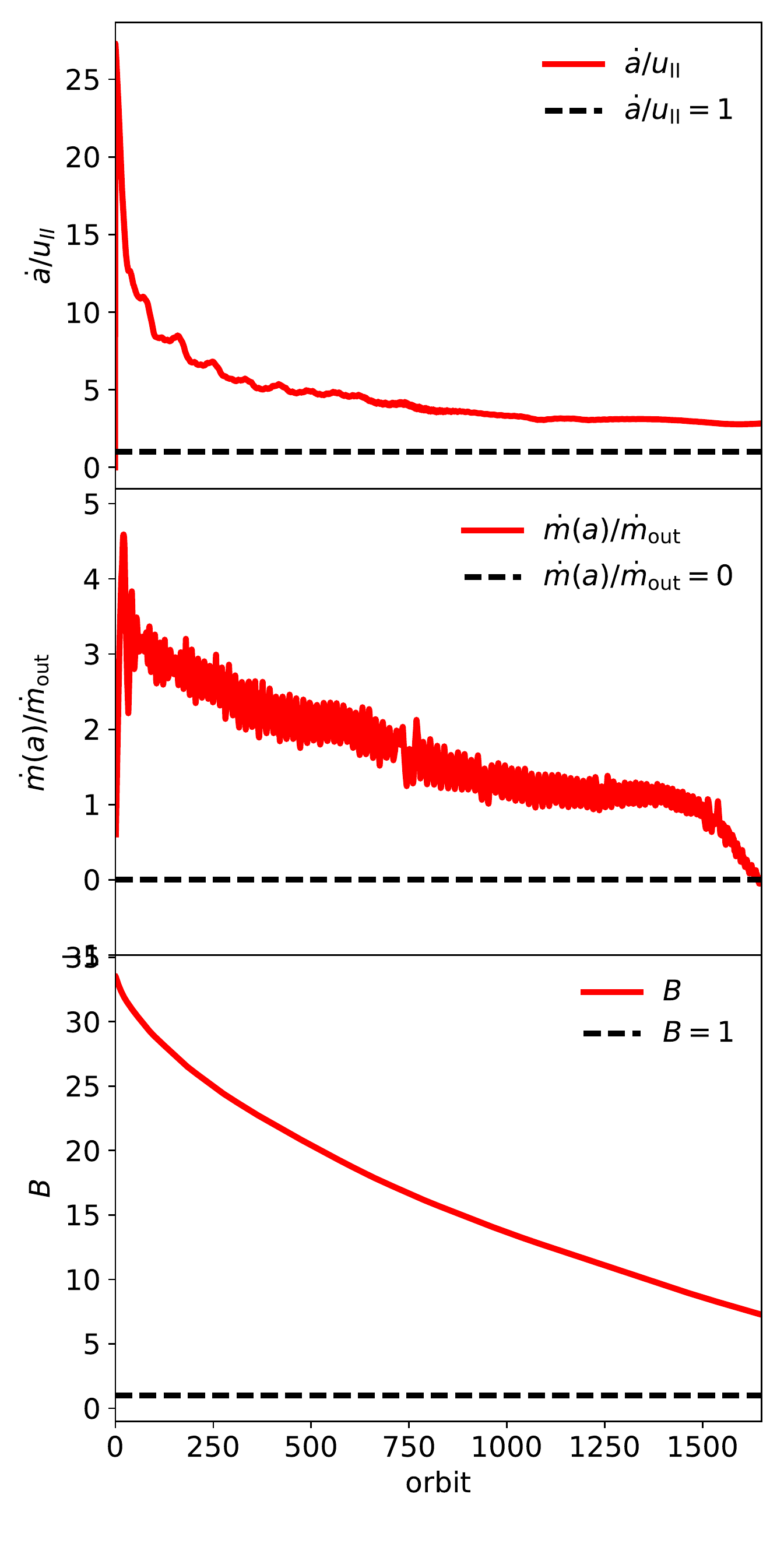}
    \caption{Standard disc. Upper panel: ratio between the planet migration velocity from simulation and its theoretical value, as a function of time (expressed in orbits) (red line), compared to the type II theoretical value (black dashed line). Middle panel: normalised accretion rate at the planet location (red line) as a function of time, compared to the theoretical expectation (black dashed line) that no mass should flow through the gap. Lower panel: parameter B as a function of orbit (red line). The disc is always more massive than the planet in this simulation, since $B>1$ (black dashed line).}
    \label{fig:h0-05_vel_mdot_B}
\end{figure}

As already observed by \citet{2015DuermannKley}, the initial planet velocity is significantly higher than the type II prediction (see the upper panel of Fig.~\ref{fig:h0-05_vel_mdot_B}); however, as we monitored the migration over a longer time (cf. $1300$ orbits in the study of \citealt{2015DuermannKley}), we can observe that the migration velocity decreases with time. Towards the end of the simulation the migration velocity has almost reached convergence (though it is still slowly declining) and is close to the theoretical value, indicating that the fast migration might be just an initial transient.

An important assumption of type II migration theory is that the disc material cannot cross the gap formed by the planet, so that the planet is locked to the disc viscous evolution. Previous studies \citep{2014Duffell,2015DuermannKley} observed that some gas does cross the gap while the planet is migrating, and this could be the cause of the planet decoupling from the gas radial evolution.
In the middle panel of Fig.~\ref{fig:h0-05_vel_mdot_B} we show the mass flow computed at planet location (i.e. the mass flow through the gap) as a function of time. While on the one hand we confirm that some gas crosses the gap, on the other hand we observe that the mass flow decreases with time and the system tends towards the prediction of zero mass flow through the gap.
It is worth underlining that the gas crossing the gap is moving from the inner part to the outer part of the disc, as $\dot{m}$ is positive; this means that the gas in the gap is \textit{acquiring} angular momentum. Note that this behaviour is the opposite of that shown in Fig.~\ref{fig:preparazione} when we did not allow the planet to migrate; in that case the mass flow was directed inwards, i.e. the gas was \textit{losing} angular momentum. Furthermore, by observing that both the mass flow and the planet velocity tend towards the type II theory prediction, it could be presumed that the mass flow is responsible for the planet fast migration; in fact, the angular momentum acquired by the gas corresponds to a tidal negative torque exerted on the planet. We will discuss this hypothesis in Section~\ref{sec: Torques on the planet}.

In the lower panel of Fig.~\ref{fig:h0-05_vel_mdot_B} we compared the evolution of parameter $B$ during the simulation (red line) to the limit value $B=1$ (black dashed line). The plot confirms that the system is in the disc-dominated regime over all the system evolution. More quantitatively, when the planet reaches the inner grid edge (approximately at $1650$ orbits) $B\sim 7$.

\section{Changing the value of $H/R$}
\label{sec: Changing the value of H/R}
\begin{figure*}
    \centering
    \includegraphics[width=1\textwidth]{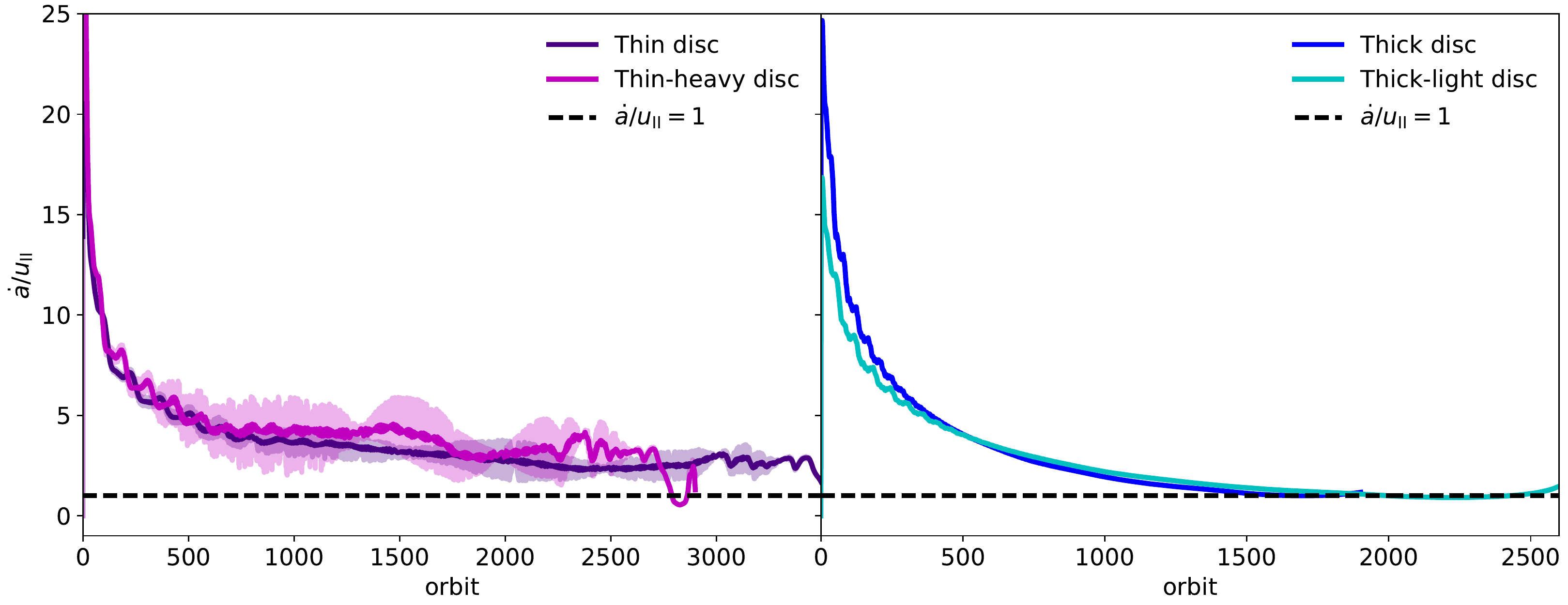}
    \caption{Left panel: ratio between simulated and theoretical velocity in simulations with $h_0=0.04$ (violet and magenta lines corresponds to the thin disc and the thin-heavy disc, respectively); the translucent lines are the raw outputs while the solid lines have been smoothed); the violet and magenta solid lines represent the smoothed velocities. Right panel: ratio between simulated and theoretical velocity in simulations with $h_0=0.06$ (blue and cyan lines corresponds to simulation the thick disc and the thick-light disc, respectively). In both the plots, the black dashed line shows the theoretical ratio.}
    \label{fig:h0-04_0-06_vel}
\end{figure*}
The standard model confirmed the findings of \citet{2015DuermannKley} that the initial migration rate can be several times higher than the type II rate, but also hinted at the fact that on long timescale the migration rate seems to converge at the theoretical value. To make further progress in understanding this convergence, we now focus on how the migration properties are affected by the change in the disc aspect ratio.
In this section we consider both higher and lower values of the aspect ratio with respect to the standard case ($h_0=0.04$ and $h_0=0.06$, respectively).

Changing the aspect ratio $h_0$ modifies also the viscosity:
\begin{equation}
    \nu \propto h_0^2,
    \label{eq:nuh0}
\end{equation}
as well as the accretion rate through the disc:
\begin{equation}
    \dot{m}\propto h_0^2\sigma_0,
\end{equation}
though in this case we can compensate for the change by changing $\sigma_0$. We performed two simulations for each value of $h_0$: in one simulation we kept the density at the same value as the standard case, thus obtaining a different value for the accretion rate; in the other simulation we adjusted the density in order to obtain the same accretion rate through the disc as the standard case. As the former simulations only differ from the standard disc due to their thickness, we named them "thin disc" (for $h_0=0.04$) and "thick disc" (for $h_0=0.06$). The latter simulations are characterised by higher and lower mass with respect to the standard disc, therefore we call them "thin-heavy disc" (for $h_0=0.04$) and "thick-light disc" (for $h_0=0.06$).

The comparison between simulations with the same aspect ratio but different density normalization is helpful in understanding the mechanism driving migration. According to the classical theory of type II migration, planet migration should be driven by viscosity, which is the same in simulation "thin disc" and in simulation "thin-heavy disc" (or, equivalently, in "thick disc" and "thick-light disc"), as $h_0$ is fixed: if classic type II is correct, we expect the two simulations to show the same qualitative migration behaviour. On the contrary, new proposed theories for giant planet migration (e.g. \citealt{2015DuermannKley}) suggest that planets are driven by density in their inward motion. Consequently, if these new pictures of migration are correct, migration in simulation "thin disc" (or "thick disc") should be different from migration in simulation "thin-heavy disc" (or "thick-light disc").
In Fig.~\ref{fig:h0-04_0-06_vel} we show the planet velocity compared to the theoretical one for both choices of $h_0$ ($h_0=0.04$ on the left panel\footnote{As the velocity in simulations with $h_0=0.04$ presents some fluctuations (violet and magenta translucent lines), we also show the smoothed velocity (violet and magenta solid lines).} and $h_0=0.06$ on the right panel), and we observe that in both cases the simulations with the same aspect ratio exhibit the same qualitative and quantitative behaviour.
This suggests that planet migration is driven by viscosity, as predicted by classical type II migration theory.
We finally note that, as in the standard setup, the velocity decreases rapidly over the first $1000$ orbits, and then it experiences a slight downward trend, tending towards the prediction of $\dot{a}/u_{\rm II}=1$.

Similarly, simulations characterised by $h_0=0.06$ begin with values of the parameter $\dot{a}/u_{\rm II}$ significantly higher than the type II prediction; at later times, however, both the the thick disc and the thick-light disc reach the predicted value $\dot{a}/u_{\rm II}=1$.
Regarding this result, it is worth observing that these two simulations are characterised by a faster viscous timescale $t_{\nu}$ with respect to the cases with smaller $h_0$, because
\begin{equation}
    t_{\nu}\propto \nu^{-1}\propto h_0^{-2}.
\end{equation}
Thus, we hypothesise that the planet in these simulations matched the theoretical prediction because the system is more evolved than the other simulated systems, having already overcome the initial transient. We will discuss further this hypothesis in section \ref{sec: Discussion}.

Note that in both the thin and the thick disc, the planet reaches the inner edge of simulation later with respect to the standard simulation. In the former case, this is a natural consequence of having a longer viscous timescale; in the latter case, it is just due to the fact that in this simulation we used a smaller $r_{\rm in}$ (see Table~\ref{tab:parameters}).

\begin{figure}
    \centering
    \includegraphics[width=1\linewidth]{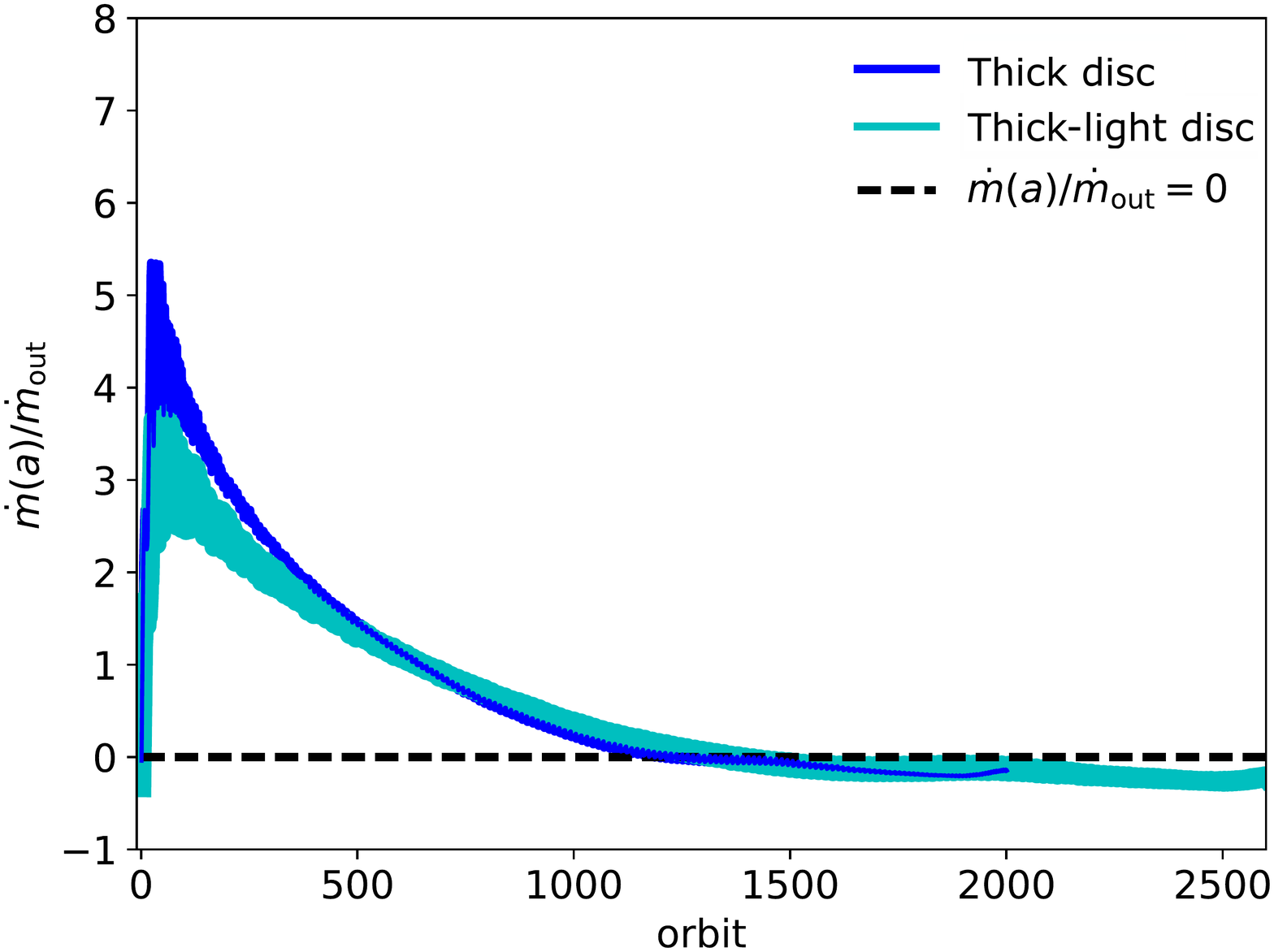}
    \caption{Mass flow through the gap for simulations with $h_0=0.06$. The blue line refers to the thick disc, the cyan line to the thick-light disc; the black dashed line illustrates the theoretical mass flow.}
    \label{fig:h0-06_mdot}
\end{figure}

As in the thick case the velocity converges towards the type II prediction, it is worth checking whether also the prediction on the gas flow across the gap is satisfied.
Therefore, we computed the normalised mass flow across the gap as a function of time (see Fig.~\ref{fig:h0-06_mdot}). When the planet reaches the condition $\dot{a}/u_{\rm II}=1$ (approximately at $1500$ orbits), the mass flow through the gap tends towards $0$.
This proves that the theoretical assumption behind classical type II migration, that there should not be mass flow across the gap, is verified.

\begin{figure*}
    \centering
    \includegraphics[width=1\linewidth]{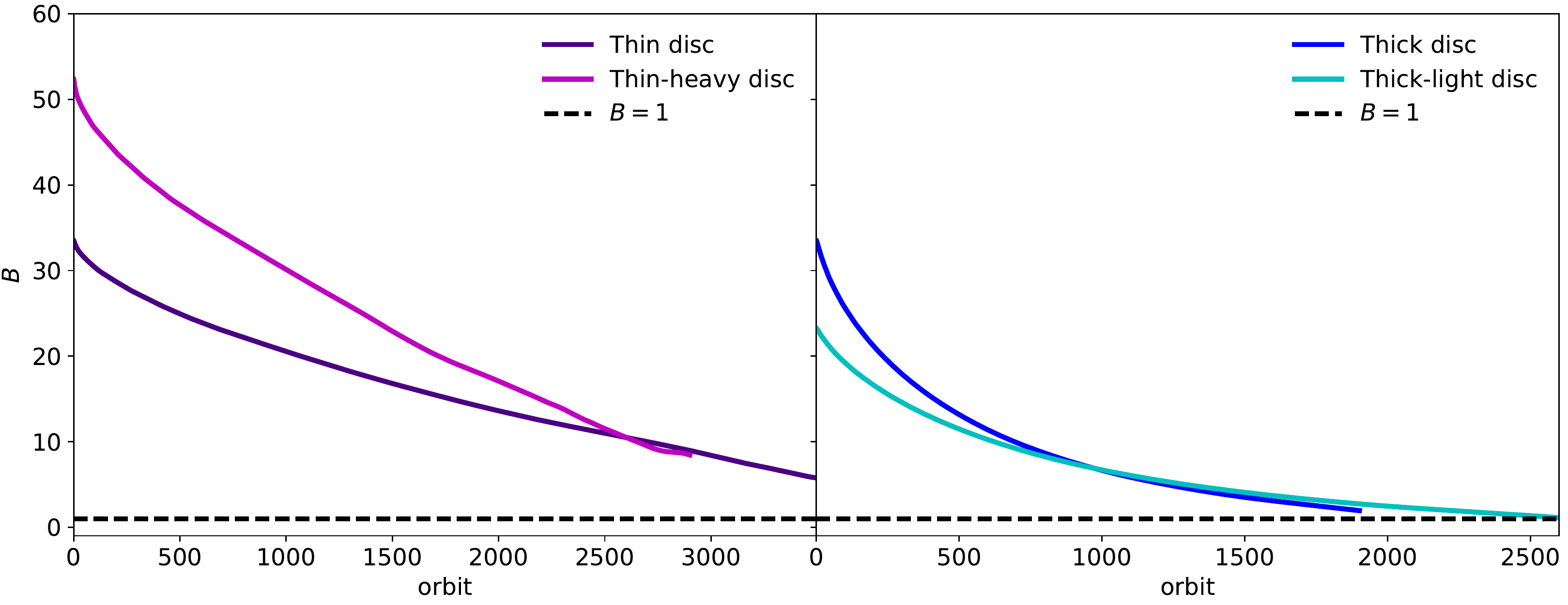}%
    \caption{Left panel: parameter $B$ as a function of the orbit number for simulations with $h_0=0.04$; the violet and the magenta lines represent the thin disc and the thin-heavy disc, respectively. Right panel: same quantity for simulations with $h_0=0.06$; the blue and the cyan lines represent the thick disc and the thick-light, respectively. The black dashed lines in both the plots show the limit between the disc-dominated and the planet-dominated regime.}
    \label{fig:h0-04_0-06_B}
\end{figure*}
Finally, to test whether the disc-dominated regime condition is satisfied, we computed the disc to planet mass ratio.
In Fig.~\ref{fig:h0-04_0-06_B} we show the variation of parameter $B$ as a function of the orbit number, considering both $h_0=0.04$ (left panel) and $h_0=0.06$ (right panel). The former case satisfies the disc-dominated condition over all the simulation. In the latter case, even if the condition $B>1$ is satisfied over all the simulation, we cannot consider $B\gg 1$ in the last orbits.
Specifically, at $1500$ orbits, when the velocity matches the prediction of type II migration, we have $B\sim 5$, and at the end of the simulation $B\sim 1$. Thus, even though it is true that the system has evolved to the type II prediction, we should make the caveat that this case is not fully in the disc dominated regime.

\section{Discussion}
\label{sec: Discussion}

\subsection{Torques on the planet}
\label{sec: Torques on the planet}
\begin{figure*}
    \centering
    \includegraphics[width=1\linewidth]{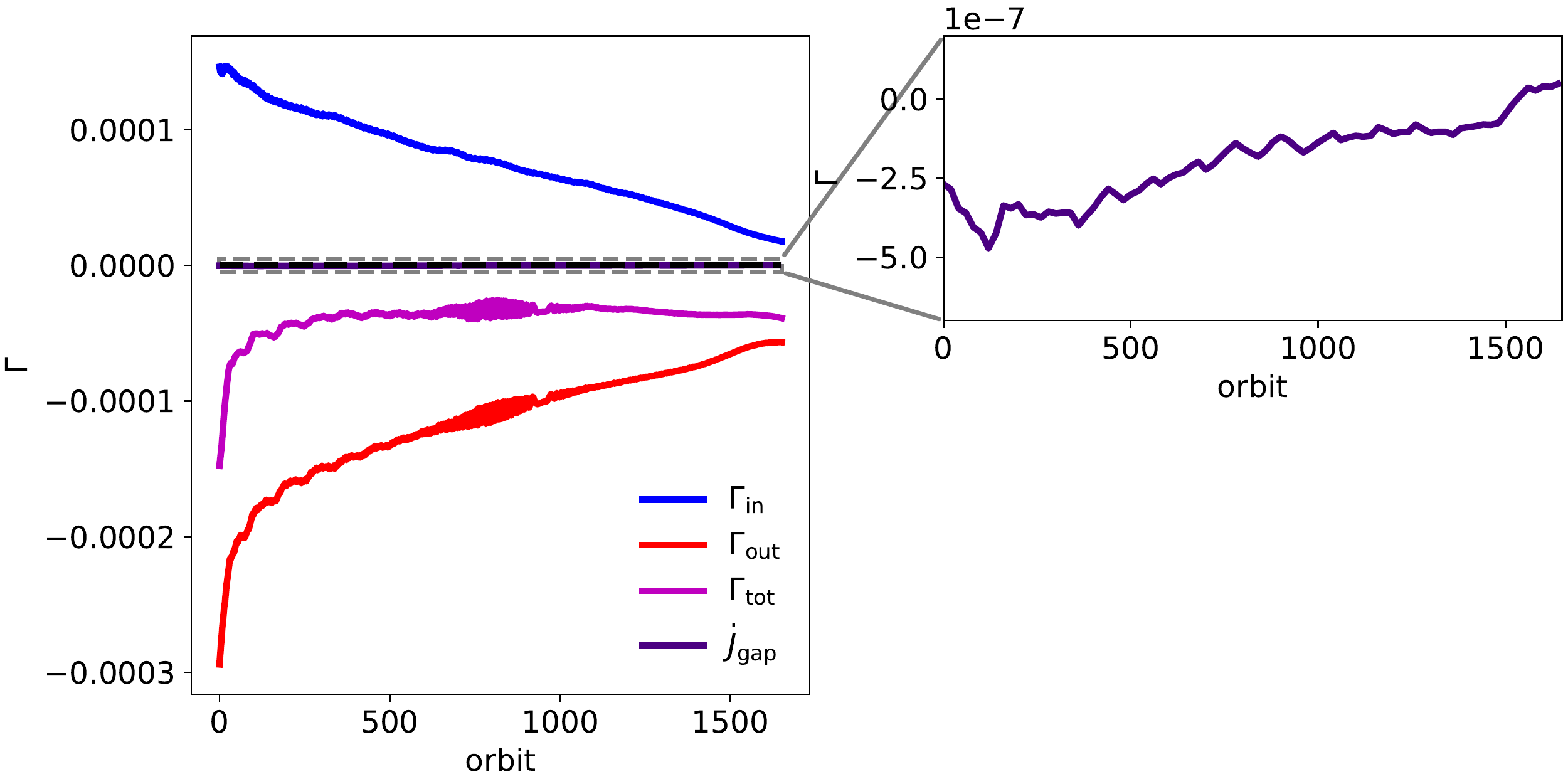}
    \caption{Total torque on the planet as a function of the orbit, for the standard disc. Blue and red lines represent the torque exerted on the planet by the disc material inner and outer to planet orbit, respectively. The magenta line represents the total torque acting on the planet. The torque exerted on the planet by the gas crossing the gap is shown in violet.}
    \label{fig:h0-05_torque_inout}
\end{figure*}

To understand why the planet slows down after the initial transient, we computed the torque acting on the planet due to the disc-planet interaction, decomposing it in the contributions from the inner and the outer disc (with respect to the planet location). As we can clearly see from Fig.~\ref{fig:h0-05_torque_inout}, the planet is subject to a negative total torque (magenta line), and thus it loses angular momentum and is pushed inwards. As it is well known in the literature about planet migration, there is a mismatch between the torques exerted by the inner (blue line) and the outer (red line) disc material, resulting in a net negative torque. We also observe that the torque becomes less negative with time, causing the planet migration velocity to slow down. In principle, there are different scenarios that could cause the reduction in the absolute value of the net torque and therefore the slowing down of migration; for example, an increase in the inner torque, at a constant outer torque. Instead, both torques are decreasing in absolute value, meaning that, in order to understand the slowing down of migration, we can focus on the reduction of the outer torque (since it is the biggest of the two).

\begin{figure}
    \centering
    \includegraphics[width=1\linewidth]{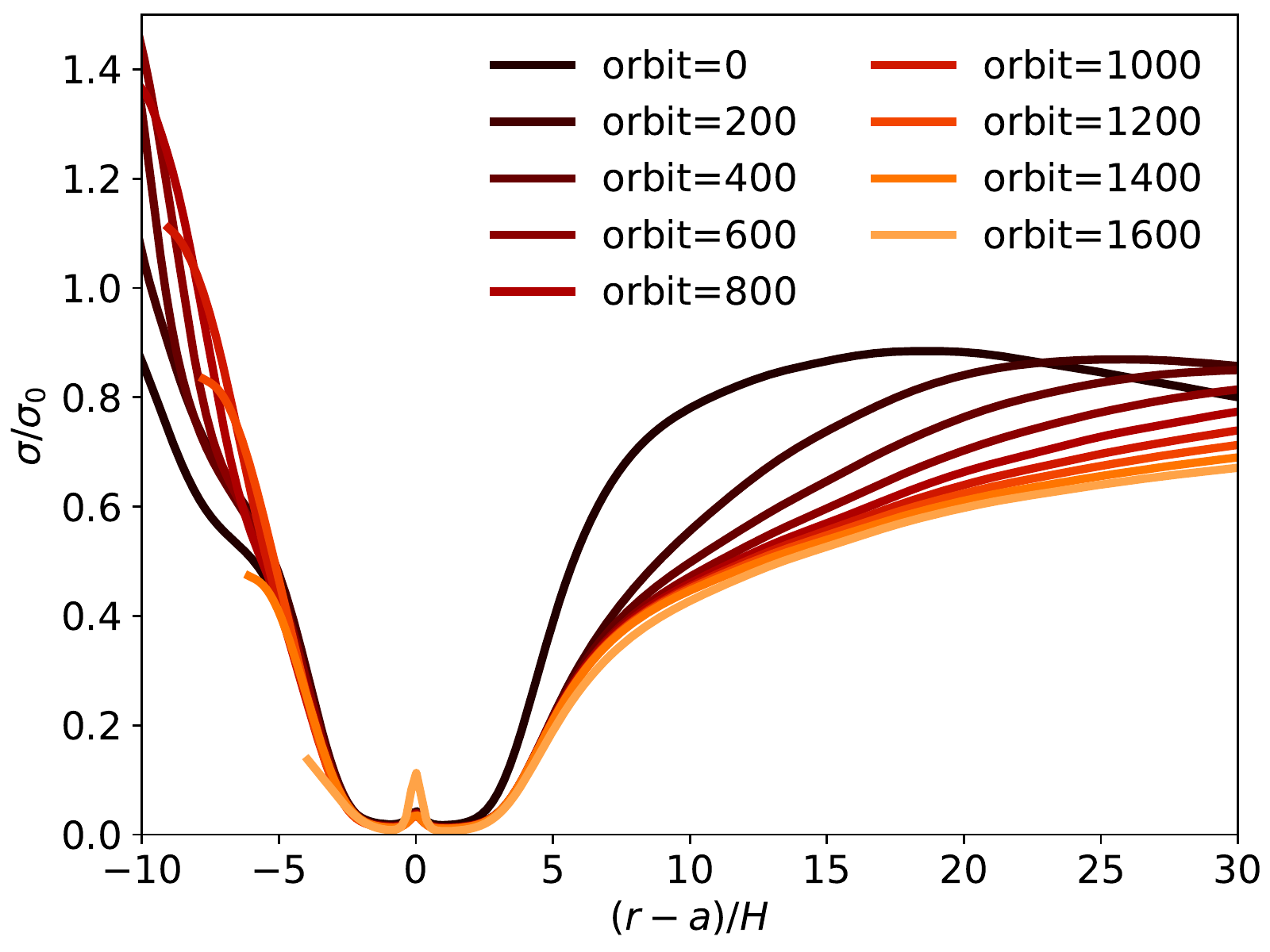}
    \caption{Gap profiles during the evolution of the standard disc. The profiles are rescaled for the disc thickness $H$ at planet position.}
    \label{fig:h0-05_dens}
\end{figure}

To understand why the outer torque is becoming smaller, we analyse the profiles of the surface density in the proximity of the gap during planet migration. These profiles (see Fig.~\ref{fig:h0-05_dens}) allowed us to observe that the gap shape changes during the system evolution. We remark that, after the planet release, the gas lags behind the planet; in fact, the planet is moving faster than the viscous inwards motion of the gas. Thus the amount of outer gas next to the planet reduces with time, causing the depletion in $\Gamma_{\rm out}$ observed in Fig.~\ref{fig:h0-05_torque_inout} and, therefore, the reduction in the planet migration velocity, observed in Fig.~\ref{fig:h0-05_vel_mdot_B}.


This behaviour, with the gap shape changing as time progresses, is somewhat reminiscent of the ``inertial limit'' \citep{Ward1997,Rafikov2002} in type I migration. In that case, as the planet migrates, it creates a ``bump'' in the surface density in the inner disc, which exerts a positive torque and slows down migration. This behaviour has been verified in the numerical simulations of \citet{Li2009} and \citet{Yu2010} (see e.g. their Fig. 4), in which the migration rate slows down considerably below the type I prediction. Note however that the behaviour we see here is slightly different: as we showed in Fig.~\ref{fig:h0-05_torque_inout}, it is the reduction of the torque coming from the outer disc, rather than the increase in the torque of the inner disc, that slows down migration.

In Section~\ref{sec: Standard setup analysis} we observed that the material crossing the gap moves from the inner to the outer disc and therefore it acquires angular momentum. As a consequence, the tidal interaction with this gas causes the planet to lose angular momentum, and it contributes to the fast planet migration. To verify whether this is a noteworthy contribution, we computed the amount of angular momentum $\dot{j}_{\rm gap}$ exerted on the planet by the material which is crossing the gap, where the gap is defined as the portion of disc next to the planet representing the $10\%$ of the total disc mass. We found that the negative torque on the planet due to the interaction with the outgoing material is 3 orders of magnitude less than the negative torque driving planet migration observed in Fig.~\ref{fig:h0-05_torque_inout} (see inset); we, therefore, conclude that this material is not directly responsible for the planet fast migration. Nevertheless, the outflowing material does modify the disc density profile, causing a variation in the gap shape, which results in a variation of the total torque; thus, this gas plays an indirect role in speeding up the planet's inward velocity.

\subsection{Comparison for different viscous time-scales}
\label{sec:comparison}

\begin{figure}
    \centering
    \includegraphics[width=1\linewidth]{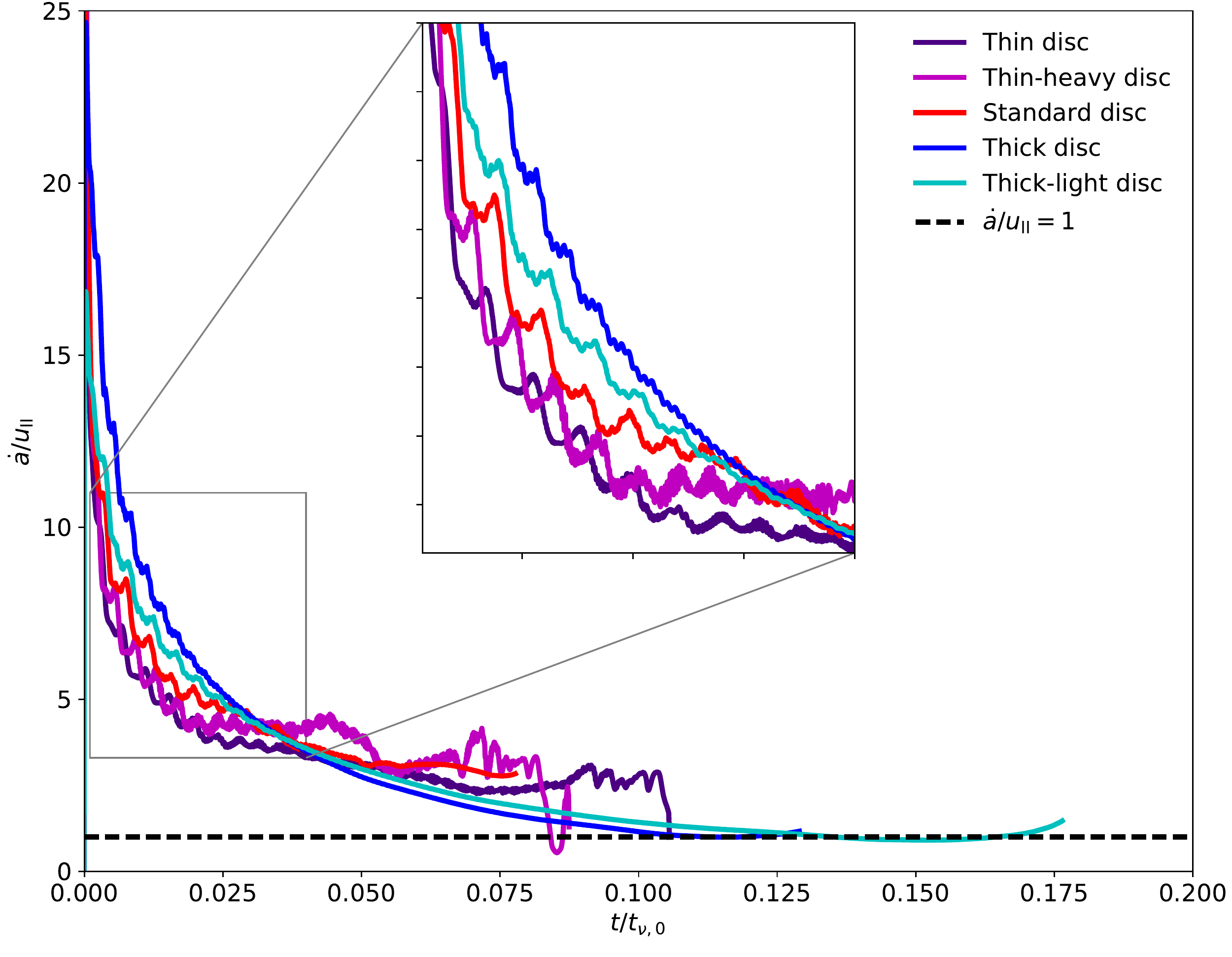}
    \caption{Ratio between simulated and theoretical planet velocity as a function of the physical time re-scaled for the viscous timescale. The colours refer to different simulations (see the legend); the black dashed line shows the theoretical value $\dot{a}/u_{\rm II}=1$.}
    \label{fig:vel_tnu}
\end{figure}

In Fig.~\ref{fig:vel_tnu} we collect all the computations of $\dot{a}/u_{\rm II}$, for comparison purposes. As systems with different $h_0$ have different viscous timescales, in order to compare them at the same evolutionary stage we need to re-scale the physical time with the viscous timescale. Therefore, we do not show $\dot{a}/u_{\rm II}$ as a function of orbits, but as a function of $t/t_{\nu,0}$, where $t$ is the physical time and $t_{\nu,0}$ is
\begin{equation}
    t_{\nu,0}=\frac{a_0^2}{\nu(a_0)},
\end{equation}
i.e. the viscous timescale computed at initial planet location.

First of all, we observe that the curves computed for different aspect ratios almost overlap when we plot them at the same viscous evolutionary stage; this proves that viscosity is the dominant process regulating the time evolution of the migration rate, confirming the hypothesis made in section~\ref{sec: Changing the value of H/R}.

Secondly, it should be noted that all the simulations have the same qualitative behaviour, characterised by a first part in which the velocity plummets from a very high value to a value of approximately $4\ u_{\rm II}$; in the second part, the evolution changes from a rapidly decreasing trend to a very slow decrease. This trend continues until the planet reaches the inner edge of the grid. Therefore, this plot suggests that all the simulations present an initial transient, that is eventually overcome if the system is given enough time to evolve. Previous claims that type II migration is not a correct description of giant planet migration focused instead on the initial transient, rather than on the long-term value. Indeed, our simulations were performed over a time period longer than preceding studies (\citealt{2015DuermannKley} stopped at around $t/t_{\nu,0}\sim 0.06$; the longest simulation we run here was evolved for a factor of $\sim$3 longer). It is precisely the increase in simulation time that allowed us to observe the tendency to evolve towards a slower migration velocity. Increasing $h_0$ here is particularly beneficial due to the reduction in viscous time scale, which allows us for the same computational time to explore a more advanced evolutionary stage.

Finally, focusing on the first part of this plot, it is interesting to observe that the simulations seem to be sorted by the aspect ratio, from the highest to the lowest value of $h_0$. The relevance of this detail consists in the fact that the thinner is the disc, the less is the amount of material expected to cross the gap \citep{Ragusa2016}. Therefore, at the beginning of migration we expect the systems with smaller $h_0$ to be more similar to the prediction rather than those with higher $h_0$. Consequently, the fact that the simulations are sorted by $h_0$ is consistent with this expectation. This also suggests that the initial transient depends on the initial conditions.

\subsection{Comparison with previous work: type II migration is correct after all}

Several papers recently revisited the issue of type II migration, as we already mentioned in the introduction. The first to raise the issue were \citet{2014Duffell}, who used a new technique in which the planet was dragged through the disc at a fixed migration rate. The migration rate was varied until a good match was found with the torque exerted by the disc onto the planet. Soon after, \citet{2015DuermannKley} confirmed these results using the more conventional technique of letting the planet free to migrate. Since both papers found a migration rate that is faster than the type II rate, this is a strong indication that the result is robust irrespective of the employed method.

Although in general the migration rate was found to be different from the type II one, note that the dependence of the migration rate on the disc mass found from these authors (see fig 3 of \citealt{2014Duffell} and fig 15 of \citealt{2015DuermannKley}) is still very similar to that predicted by type II theory: it rises approximately linearly with low-mass discs, saturating when the disc becomes more massive than the planet. More recently, \citet{Kanagawa2018} also found very similar results, see their figure 6.

The work of \citet{Kanagawa2018} deserves some further discussion because these authors claimed that the migration rate of massive planets is an extension of type I migration rate, but with a depleted surface density in the co-orbital region. Our calculations allow us to reconsider this conclusion. As we showed in section \ref{sec: Torques on the planet}, in fact the torque on the planet coming from material in the co-orbital region is negligible. The depletion in this region means that the torque is dominated by material at the edge of the gap and therefore type II migration is intrinsically different from type I. Intriguingly, \citet{Kanagawa2018} do find that the migration rate correlates with the surface density at the bottom of the gap (though with a large scatter). However, the existence of a correlation does not prove a causal connection, namely that the surface density at the bottom of the gap drives migration. The correlation can also be explained if both quantities are set by a third quantity, in this case the viscous flux of angular momentum (see section 2.2 of \citealt{Kanagawa2018}).

The closest work to the one we present here is \citet{2018Robert}. These authors showed that the migration rate scales with the viscosity, as expected in type II migration. However, the rate never assumes the viscous value (see their fig. 2) and the planet eventually migrates \textit{slower} than viscous. We remark that these authors did not consider the variation of the parameter B (equation \ref{eq:B}) as the planet migrates; therefore, the fact that the planet migrates more slowly than viscous does not mean that the migration rate is not compatible with type II. These authors also considered the interesting case of a disc which is set up to have a zero net mass accretion rate (their case B1). At odds with type II migration theory, predicting that the planet should not migrate, the migration rate is in fact fairly similar to the standard case of a disc with a net mass inflow (their case A1). Note however that, although such a disc is formally in steady state because the mass flux is null, as soon as perturbations (such as those induced by a planet) are introduced, the surface density profile no longer satisfies the relation of steady state accretion discs $\dot{M} \propto \nu \Sigma$. Therefore, these perturbations will tend to modify the surface density profile towards the situation in which $\nu \Sigma$ is a constant, i.e. the initial conditions of case A1. This possibly explains why their two cases A1 and B1 have a fairly similar migration history.

The analysis we present in this paper allows us to explain simultaneously all the relevant simulations on this subject in the last years. Indeed, the initial migration rate is much higher than type II. However, with enough time the migration rate does converge to the type II value, as we showed in section \ref{sec:comparison}; this result had been missed before because the simulations had not been run for sufficient time (compared to the viscous timescale). As was already highlighted by those previous works, the scaling of the migration rate with viscosity and planet to disc mass ratio was already shown to scale in the same way as the type II rate; our work confirms that also the normalization constant is eventually the same.

From our simulations we can also understand why the initial transient cannot be maintained. As we show in fig \ref{fig:h0-05_dens}, the gas is ``left behind'' because it cannot move at a rate faster than viscous. The result is that less material is present in the vicinity of the planet outside its orbit and migration slows down. An equilibrium can be reached only when the planet moves at the same rate as the gas. Our simulations thus prove that, as was traditionally assumed in the type II picture, the surface density of the disc eventually re-arranges itself to move the planet at the viscous rate.

We can thus conclude that the classical picture of type II migration is actually correct, despite the recent claims on this topic in the last years. There is however a new aspect of the topic, that was only uncovered by these recent simulations. Namely, converging to the type II rate takes a sizeable fraction of the viscous time-scale. This time-scale is potentially long and therefore the transient is important.

Lastly, it should be noted that, although mass transport through the gap does happen, it is not the reason for the apparent discrepancy from type II migration. \citet{2018Robert} showed that one obtains similar migration rates also when cutting the mass transport through the gap. Along the same lines, our simulations show that the mass transport through the gap decreases with time and eventually becomes smaller than the viscous rate.

\subsection{Limitations and further observations}
Through this work we showed that, in contrast to previous claims, planet migration tends towards the classical type II migration regime. However, it should be underlined that the convergence to type II migration requires a significant fraction of the viscous timescale, and, during this transient, the planet experiences a non-negligible inward migration. As an example, in the system with $h_0=0.06$ the transient lasts approximately $t=0.1\ t_{\nu}$. During this time, the planet has migrated from $r=1$ to $r=0.2$, a factor of 5 in radius. Therefore, the fast-migration transient might still be physically relevant even if it is not the long term value.

It is worth stressing however that, in order to start from a steady state initial condition for the planet-disc system, the initial conditions we have assumed in this work are highly artificial: we fixed the planet at its initial location for $7000$ orbits. In a more realistic situation, the planet would start from a lower mass and then migrate while increasing its mass, until becoming massive enough to open a gap. It is not clear how the transient would change taking into account a realistic increase in planet mass, but we can speculate that this would reduce the importance of the transient since the surface density profile would be closer to the quasi-steady state for the instantaneous planet mass.

Even though the simulations presented in this paper are among the longest ever presented in the literature for the case of type II migration\footnote{The longest simulation exhibiting type II migration is \cite{2018Ragusa}}, our longest simulation lasts for $t\sim 0.175\ t_{\nu}$, which is only a fraction of the viscous timescale. Since type II migration relies on a viscous process, further understanding of type II migration requires to run these simulations for a time comparable with the viscous timescale.

The simulations presented in this work show another feature that is worth to be highlighted. In fact, in these simulations, we observed a significant modification in the planet eccentricity and we found that the eccentricity increases more in the thin discs rather than in the thick ones. Moreover, we noticed that a slight modification of the aspect ratio can produce a substantial difference in the eccentricity outcome. Although the analysis of this aspect is beyond the purpose of the present work, it may be interesting to investigate it further in future works.

Finally, our simulations present the same limitations as other works in the literature: in this work we simply assumed an ideal $\alpha$ viscosity, characterised by a constant $\alpha$ over all the disc, and two dimensional, isothermal discs. Since the migration is mostly driven by material in the midplane and proto-planetary discs usually have short cooling times (at least in their outer parts), the last two assumptions are reasonably well-motivated. Given the current debates about the origin of angular momentum transport, it is not clear instead how representative the ideal $\alpha$ case is; for example, discs might be almost inviscid (with accretion driven by MHD winds), or even if viscous, the turbulence could be highly anisotropic. The migration rates in these cases would be severely affected \citep{McNally2019}. What is clear, though, is that the recent progresses in type II migration theory are forcing us to reconsider the basic understanding even of the idealised case. Certainly no progress can be made for the more complicated cases if even the idealised case is not well understood.

\section{Conclusions}
\label{sec: Conclusions}
In this paper we run long-term simulations of massive (gap-opening) planets. Thanks to the extended time span we simulated, we obtained evidence that the fast migration found in previous studies is just an initial transient and not the actual planet migration rate. Firstly, we found that even if it is true that the planet migrates faster than that predicted by the type II theory, it is also true that planet velocity decreases with time and it continues to slightly decrease at later orbits; if velocity keeps that downward trend, it will finally reach the theoretical value. Secondly, we confirmed that some gas can cross the gap formed by the planet in the disc surface density, but we also showed that the mass flow through the gap experiences a decrease during the system evolution; if, in the end, gas stops crossing the gap, then the planet will be locked to the viscous evolution. 
Thirdly, we analysed the torques acting on the planet, and we found that it is experiencing a negative torque which decreases significantly in absolute value with time - a result of the gas lagging behind the planet, because the gas cannot keep up with the faster than viscous motion of the planet, implying that such a fast migration rate cannot be sustained for a long time. As this behaviour corresponds to a slowing down of the migration, the torques tends towards the value required to obtain the velocity predicted by type II migration.

Motivated by these observations, we explored in detail the dependence of the results on the aspect ratio $H/R$. The reason why this investigation is important is that when we change $H/R$, we are also varying (with respect to the standard case) either the value of the disc density or the value of mass flow rate through the disc. By renormalising the migration rate to the viscous time, we notice that all the simulations we run exhibit a very similar time behaviour. This is a strong indication that viscosity is the driver of the migration process (as suggested by the classical theory of type II migration), rather than the disc density (as suggested by the new studies of giant planet migration). In the case of the simulation with the higher viscosity we run, therefore the shorter viscous time-scale, we also noticed that the migration rate eventually reaches the type II rate.

Therefore, our results suggest that type II migration is a correct paradigm after all. However, reaching the type II migration rate requires a potentially long time, a sizeable fraction of the viscous time-scale; the initial transient might therefore be very important to describe accurately the planet motion. Ultimately, because the simulation we have run here employ highly artificial initial conditions, the way forward is to allow the planet to grow at realistic rates while it migrates.

Finally, although we show that the planet velocity tends towards the predicted type II rate at late times, there is no guarantee that the planet will ever attain the viscous speed before it enters the planet dominated regime (i.e. $B \sim 1 $, \autoref{eq:B}): indeed our simulations with $h=0.06$ (which has run for the largest fraction of a viscous time) is already entering the regime with $B \sim 1$ at the end of the simulation.

\section*{Acknowledgments}
We thank the anonymous referee for his report that improved the clarity of the paper.
This project has received funding from the European Union’s Horizon 2020 research and innovation programme under the Marie Skłodowska-Curie grant agreement No 823823 (DUSTBUSTERS). This work has been supported by the DISCSIM project, grant agreement 341137 funded by the European Research Council under ERC-2013-ADG. This work is part of the research programme VENI with project number 016.Veni.192.233, which is (partly) financed by the Dutch Research Council (NWO).

\bibliographystyle{mnras}
\bibliography{biblio}

\bsp	
\label{lastpage}
\end{document}